\documentclass[twocolumn,showpacs,floatfix,superscriptaddress,citeautoscript,cite]{revtex4}
\usepackage{graphicx}
\usepackage{xfrac}

\begin{document}

	\author{J. Sivek}
		\email{jozef.sivek@ua.ac.be}
	\author{H. Sahin}
		\email{hasan.sahin@ua.ac.be}
	\author{B. Partoens}
		\email{bart.partoens@ua.ac.be}
	\author{F. M. Peeters}
		\email{francois.peeters@ua.ac.be}

	\affiliation{Departement Fysica, Universiteit Antwerpen, Groenenborgerlaan 171, B-2020 Antwerpen, Belgium}

	\title{Adsorption and absorption of Boron, Nitrogen, Aluminium and Phosphorus on Silicene: stability, electronic and phonon properties}

	\date{\today}


\begin{abstract}
\textit{Ab initio} calculations within the density-functional theory formalism are performed to investigate the chemical functionalization of a graphene-like monolayer of silicon -- silicene -- with B, N, Al or P atoms. The structural, electronic, magnetic and vibrational properties are reported. The most preferable adsorption sites are found to be valley, bridge, valley and hill site for B, N, Al and P adatoms, respectively. All the relaxed systems with adsorbed/substituted atoms exhibit metallic behaviour with strongly bonded B, N, Al, and P atoms accompanied by an appreciable electron transfer from silicene to the B, N and P adatom/substituent. The Al atoms exhibit opposite charge transfer, with n-type doping of silicene and weaker bonding. The adatoms/substituents induce characteristic branches in the phonon spectrum of silicene, which can be probed by Raman measurements. Using molecular dynamics we found that the systems under study are stable up to at least T = 500 K. Our results demonstrate that silicene has a very reactive and functionalizable surface.

\end{abstract}

\maketitle

\section{Introduction}

Since the first reports on the successful isolation
of a stable monolayer of graphene \cite{novoselov_2004, geim_2007}
special efforts have been invested in the exploration
of similar materials with novel properties resulting
from their ultra-thin two-dimensional nature.
Those alternative materials have the potential
to bypass some of the obstacles existing
in the usage of graphene in contemporary electronics i.e.,
incompatibility with present day silicon technology
and lack of an energy bandgap which is essential
for all semiconductor devices. For graphene
itself further modification is inevitable
in order to tailor its electronic properties.
A broad range of chemical decoration \cite{elias_2009, nair_2010, sahin_2011, ortwin_2010-fh_graphene, sahin_2012-cl_g}
and substitutional doping \cite{panchakarla_2009,sheng_2012-g_b,wang_2010-g_n}
have been investigated to open and tune
the band gap of graphene. Particular attention was
given to boron and nitrogen as the first
choice elements because of their chemical
propinquity to carbon as well as to silicon. \cite{lherbier_2008}
The second row elements have not been
omitted, Al, Si and P doping induce band gap
opening in monolayer graphene. \cite{denis_2010-second_row_el}
Theoretical studies of functionalization
of graphene have later been supported by
the experimental realization of substitutional
doping with B and N atoms via arc discharge
using graphite electrodes in the presence
of hydrogen and B or N atom incorporating
molecules (pyridine, amonia, $\mathrm{B}_{2}\mathrm{H}_{6}$),\cite{panchakarla_2009}
thermal annealing of graphene in the presence
of boron oxide \cite{sheng_2012-g_b} or nitrogen
plasma treatment of graphene.\cite{wang_2010-g_n}

\begin{figure}
	\includegraphics[width=8.5cm]{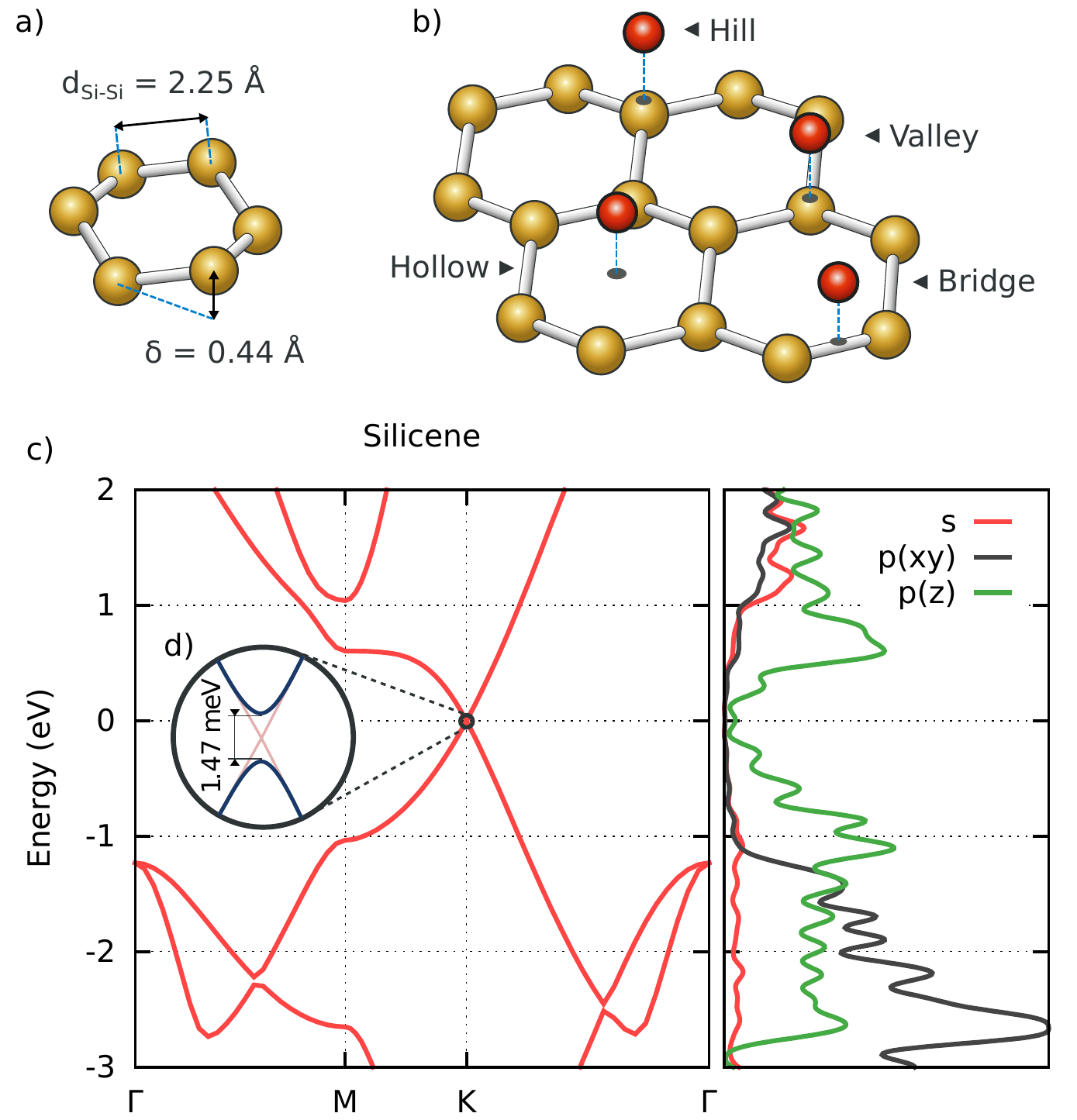}
	\caption{\label{fig-silicene}
	(Color online) (a) Structural parameters for silicene
	and (b) possible adsorption sites on the silicene lattice.
	(c) Electronic band dispersion and partial density of states
	for perfect silicene. The energies are relative to
	the Fermi level (i.e., $E_{\mathrm{F}}=0$).
	The inset (d) displays the calculated spin-orbit gap of 1.47~meV. 
	}
\end{figure}

Recently, the monolayer honeycomb structure of silicon,
silicene, has emerged as a potential few-atom-thick
material to replace graphene. As was reported
by early theoretical works, silicene is a semimetal with
linearly crossing bands and a zero electronic band gap 
similar to graphene. \cite{takeda_1994,cahangirov_2009,sahin_2009}
Furthermore, similar to graphene, electrons propagating
through the monolayer crystal structure of silicene
are predicted to show massless fermion behavior
in the vicinity of the Dirac point. Additionally,
some unique features of monolayer silicene such
as quantum spin Hall effect,\cite{liu_2011-2d_ge}
a large spin-orbit interaction,\cite{liu_2011-s-o-si-ge}
a mechanically tunable bandgap\cite{topsakal_2010-deformation_of_2d_materials}
and a valley-polarized metallic phase\cite{ezawa_2012-qhe-silicene}
have been reported by theoretical studies.
Moreover, the recent experimental observations
and synthesis of silicene \cite{padova_2010,padova_2008,vogt_2012,lin_2012,jamgotchian_2012}
have opened a new path for nanoscale materials
which might be easily functionalized chemically
or mechanically and incorporated within electronics
as we know it today.\cite{fleurence_2012}
Very recently we reported the adsorption
characteristics of alkali, alkaline earth
and transition metal atoms on monolayer silicene.\cite{sahin_2013-silicene}
Differing from graphene, one can expect
the substitution of B, N, Al or P atoms
to be more likely on the silicene surface (Fig. \ref{fig-silicene}(a)),
as it is more reactive due to its $sp^{3}$-like lattice structure.

In this paper, motivated by the route undertaken
with graphene and the recent synthesis of silicene,
we investigate the adsorption and absorption of B, N, Al and P atoms
on the surface of pristine free-standing silicene
together with their substitutional doping.
Our interest is pointed towards the compounds' structure,
binding energies of the most stable structures and their electronic,
magnetic and phonon properties. We find that
the adsorbed B, N and P atoms
are strongly bonded to the underlying silicene,
attached to its surface but also penetrated into the
silicon lattice. Weaker bonding and immersion is
observed for the Al atom. These observations
differ from the general chemical functionalization
of graphene which is highly dependent on the presence
of defects and crystal boundaries.\cite{oubal_2012}
The stability of the ground state structures
has also been addressed by using \textit{ab initio}
molecular dynamics and phonon calculations.

\section{Computational Methodology}

Our present investigation of the chemical
modification of silicene employs density functional theory (DFT)
as implemented in the VASP package.\cite{kresse_1996}
We have made use of the spin polarized local-density
approximation (LDA) \cite{ceperley_1980-lda}
for the exchange-correlation functional, the projector
augmented wave method \cite{blochl_1994-paw} and a plane-wave
basis set with an energy cutoff of 500 eV.
The sampling of the Brillouin zone was done
for the supercell with the equivalent of
a $24\times24\times1$ Monkhorst-Pack $k$-point grid
for a silicene unit cell (containing 2 silicon atoms).
The partial occupancies in the electronic
ground state calculation were treated using
the tetrahedron methodology with Bl\"ochl corrections \cite{blochl_1994-tetrahedron}.
For the purpose of the calculation of the density of states 
a Gaussian smearing of the energy levels was applied
with standard deviation set to 0.1~eV.

To eliminate the interaction emerging from periodic
boundary conditions in all three dimensions
a $4\times4$ supercell was used with the height of
15~{\AA} to include enough vacuum, and dipole corrections
were used. All reported quantitative results of the charge
transfer were obtained by usage of the Bader charge
population analysis \cite{bader_1991, henkelman_2006},
and the iterative modified Hirshfeld charge population
analysis.\ \cite{hirshfeld_1977, bultinck_2007-ih}

The relaxation of the atomic positions
was performed with forces smaller than 0.01~eV\AA$^{-1}$.
To reduce the strain induced by the adsorbates/substituents,
the lattice parameters were optimized properly.
The phonon frequencies for adatom adsorbed/substituted
silicene were calculated using
the Small Displacement Method.\cite{alfe_2009}

\textit{Ab initio} molecular dynamics simulations (MD)
were performed with use of the non-spin polarized
local-density approximation (LDA) \cite{ceperley_1980-lda}
for the exchange-correlation functional,
the projector augmented wave method \cite{blochl_1994-paw}
and a plane-wave basis set with an energy
cutoff of 500 eV. The sampling of the Brillouin zone
was done for the supercell with the equivalent of
a $12\times12\times1$ Monkhorst-Pack $k$-point grid
for a silicene unit cell. The integration of
Newton's equations of motion was performed
using the Verlet algorithm where
Harris corrections were used in order
to correct the forces. The simulations
were performed within the micro-canonical (NVE) ensemble
with velocities assigned according to
the Maxwell-Boltzmann distribution
at the temperature of 500 K during
the entire calculation. To avoid large
temperature fluctuations velocities
were normalized every 40 steps.
The total duration of the simulation
was 2 ps with the time step equal to 1 fs.

\section{Results and Discussion}

\subsection{Atomic Structure and Migration Barriers}

First, we consider the adsorption of a single
atom on a silicene surface. In contrast to the completely
flat one-atom-thick surface of graphene, silicene is buckled
as can be seen in Fig. \ref{fig-silicene}(a) and we can expect
higher reactivity due to this $sp^{3}$-like lattice structure.
Similar to graphene, silicene is a semimetal with linearly
crossing bands at the Fermi level with a zero electronic
band gap (Fig. \ref{fig-silicene}(c)). However, effect
of spin-orbit coupling, that yields 1.47~meV spin-orbit gap
(Fig. \ref{fig-silicene}(d)), is much larger than 
that of graphene's.\cite{ezawa_2012-soc_silicene}
We define the binding energy for adsorption as:
$E_{\mathrm{B}} = E_{\mathrm{s}ystem} - (E_{\mathrm{silicene}} + E_{\mathrm{adatom}})$.
There are three possible adsorption sites on graphene, while
as a consequence of the buckled hexagonal lattice structure 
of silicene has now four different adsorption sites
as shown in Fig. \ref{fig-silicene}(b):
above the center of the hexagonal silicon rings (Hollow site),
on top of the upper silicon atoms (Hill-site), on top of
the lower silicon atoms (Valley-site) and on top of
the Si-Si bond (Bridge site).

\begin{figure}
	\includegraphics[width=8.5cm]{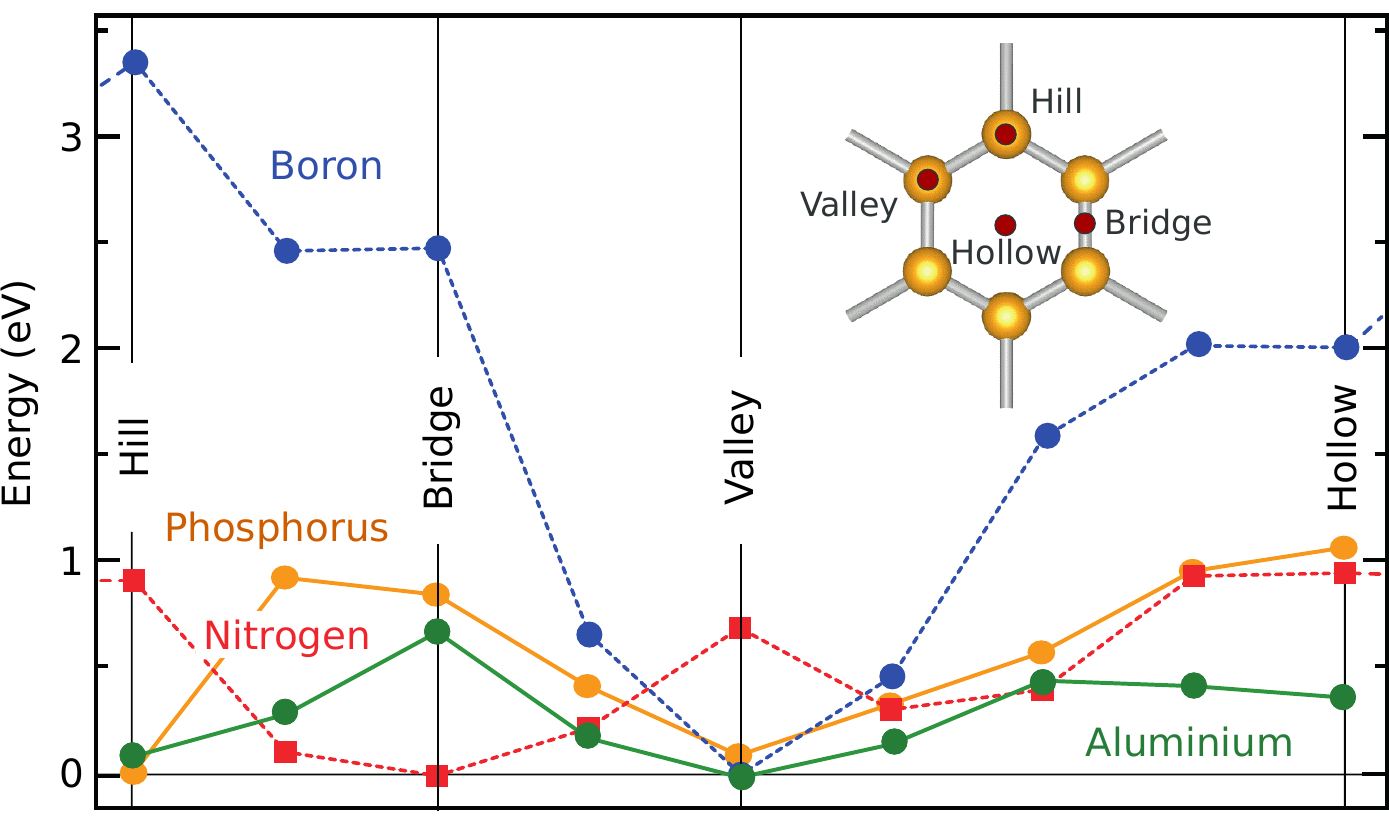}
	\caption{\label{fig2}
	(Color online) Diffusion characteristics of B, N, Al and P adsorbates.
	The energies are relative to the binding energy of the most preferable adsorption site.
	Spatial distance is plotted on x axis, the intermediate
	positions between high symmetry points were obtained by 
	restricting the adatoms movement perpendicular
	to the silicene surface. The Local minima in diffusion
	energy plots outside of the four high symmetry
	adsorption sites do not provide stable binding
	without aforementioned restrictions.
	}
\end{figure}

The possible diffusion paths were determined
from the energetics of the adatoms on the different
lattice points. As shown in Fig.~\ref{fig2},
each N adsorbate has to overcome an energy barrier
of $\sim$0.7 eV for diffusion from one bridge site
to a neighboring one via a valley site.
B adsorbates, because of their stronger binding energy
compared to N atoms, see larger migration barriers.
Diffusion of a B atom from one valley site to 
another one can occur via a hollow site by overcoming
the energy barrier of $\sim$2.0 eV.
The P adatom has to overcome an energy
barrier of $\sim$0.9 eV when diffusing
between neighboring hill sites via the bridge and valley sites.
For the Al adatoms the energy barrier is $\sim$0.4 eV
between two neighboring valley sites.
After unconstrained relaxation, the four
adsorption sites provide stable or metastable
binding, albeit with different binding energy.
Only the adsorption of an Al atom on a bridge
site is found to be unstable.

For the graphene surface
the bridge site was found to be the most favorable
adsorption site for both B and N adsorbates.\cite{nakada_2011}
However, we find that the most favorable adsorption
sites on the silicene surface are the valley
and bridge sites for B and N adsorbates, 
respectively. While the adsorption of the B (N) atom occurs
with a -1.8 (-4.6) eV binding energy on graphene,\cite{nakada_2011}
for silicene the bonding between B (N) and the silicene lattice
is stronger with a binding energy of -5.85 (-5.54) eV.
The Al and P atom preferential adsorption sites
were found to be the valley site with a binding
energy of -2.87 eV and the hill site with an adsorption
energy of -5.28 eV, respectively.
We found the adatoms to considerably distort
the underlying silicene layer.
A B adsorbate is almost completely immersed into the silicene layer
and pushes the underling Si atom down from its original position.
The B adatom average distance from the underlying Si layer
is only 0.71~{\AA} as can be seen in Table \ref{table_main}.
N atom adsorption on the bridge site results in Si-Si bond breakage.
A similar effect as with the B adsorption occurs for P and Al adatoms, albeit
with lower intensity and with up to two times larger average
distances from the underlying Si layer as can be seen in Table \ref{table_main}.

\begin{figure*}
	\includegraphics[width=17cm]{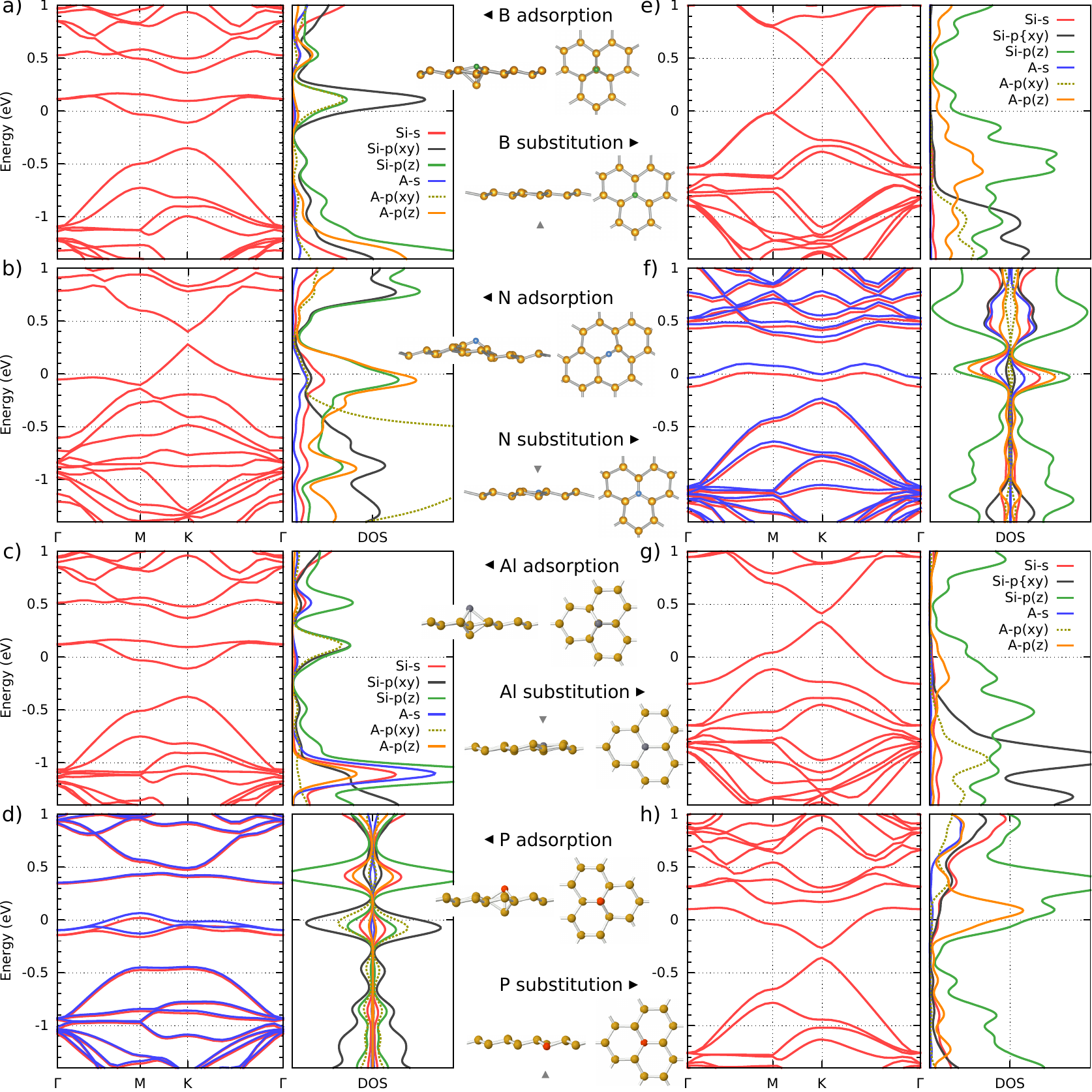}
	\caption{\label{fig3}
	(Color online) Electronic band dispersions,
	decomposed density of states (DOS) together with
	side and top view of the final relaxed structures.
	(a) Adsorption of B on valley site,
	(b) adsorption of N on bridge site,
	(c) adsorption of Al on valley site,
	(d) adsorption of P on hill site,
	(e) B, (f) N, (g) Al and (h) P substituted silicene layers.
	The energies are relative to the Fermi level ($E_{\mathrm{F}}=0$).
	In the legend ``A'' refers to the adsorbed/substituent atom.
	}
\end{figure*}

Next we investigate the absorption, i.e., the substitution of a single Si
atom with a B, N, Al or P atom. 
The binding energies for substitutional doping are
-6.21, -5.64, -2.28 and -4.84~eV, for a B, N, Al and P atom, respectively.
This binding energy for substutution is calculated as follows:
$E_{\mathrm{B}} = E_{\mathrm{system}} - (E_{\mathrm{silicene}} (N_{\mathrm{Si}} - 1)/N_{\mathrm{Si}} + E_{\mathrm{adatom}})$.
B-Si and N-Si bonds are shorter than the intrinsic
Si-Si bond lengths, the P-Si bond is also slightly
shorter but the Al-Si bond lenght is larger by 0.1~{\AA}
in comparison to the Si-Si bond length
as can be seen in Table \ref{table_main}
(comparabable to N and B substituted graphene \cite{panchakarla_2009}).
The B, N and Al substituent atom bonds with neighboring Si atoms
are more flat, reducing local buckling. The P substituent atom exhibits
comparable buckling as the Si atoms, the angle $\theta_{\mathrm{SiPSi}}$
(111.3$^{\circ}$) is even smaller compared to
$\theta_{\mathrm{SiSiSi}}$ (116.3$^{\circ}$) of pure silicene.
With increased distance from the B, N, Al or P atom
the buckling of silicene is quickly restored.
All the substituents except B atom exhibit
a small displacement from the containing Si layer.
As already mentioned, the B atom adsorption on the valley site 
(to some degree P atom adsorption on the hill as well) 
results in structures resembling substitution structures,
with the Si atom pushed out of the silicene plane and
adsorbate immersion into the plane. This can be
a potential route for chemical doping
of silicene via chemical decoration.

The calculated values for the binding energies show
that substitutional doping is energetically more favorable
for B and N atoms, whereas it is less favorable
for Al and P atoms compared to adsorption. This behavior of
the binding energy causes a lower amount of immersion
for the Al and P atoms as compared to the B and N atoms.

We find the adsorbed as well as substituent Al atom to have
notable lower absolute value for its binding energy in comparison to
the other investigated atoms. The reason for this different
behavior can be explained from a chemical point
of view. The Al atom is the only one, due to its lower
electronegativity compared to the Si atom, that looses electrons
in favor of the neighboring Si atoms. The induced
depopulation of bonding orbitals increases
the length of the Si-Al bond ($d_{\mathrm{SiAl}}$)
and weakens it.

\subsection{Electronic Structure}

\begin{table}
	\caption{\label{table_main}
	Calculated parameters for B, N, Al and P adsorbed/substituted silicene:
	lattice constant ($a$); adsorbate-Si bond
	distance ($d_{\mathrm{SiX}}$); average distance to
	low-lying Si layer ($h$); magnetic moment ($\mu$);
	binding energy per adatom/substituent($E_{\mathrm{B}}$); work function ($\Phi$,
	for the adsorbtion the value correspond to
	the value measured from the side of the adsorbate);
	charge located on adatom/substituent obtained with Bader charge analysis  ($\rho^{\mathrm{B}}$),
	or iterative Hirshfeld method ($\rho^{\mathrm{H}}$)
	and dipole moment ($p$).
	The adsorbate and substituent concentration
	was one to 32 and 31 Si atoms, respectively.
	}

\begin{tabular}{ccccccccccccccc}
	\hline  \hline
			& site &  $a$  & $d_{\mathrm{SiX}}$ & $h$  & $\mu$     & $E_{\mathrm{B}}$ & $\Phi$ & $\rho^{\mathrm{B}}$ & $\rho^{\mathrm{H}}$ &  $p$   \\ 
			&      &  \AA  & \AA                &\AA   & $\mu_{\mathrm{B}}$ & eV    & eV   & e          & e         & e$\cdot$\AA\\
	\hline \hline           
	B (ads) & V    & 15.16 & 1.95               & 0.71 &  0.0      & -5.85    & 4.73   &-1.6        &-0.7        & -0.02  \\ 
	N (ads) & B    & 15.30 & 1.63               & 1.42 &  0.0      & -5.54    & 5.02   &-2.0        &-0.6        &  0.06  \\ 
	Al (ads)& V    & 15.26 & 2.51               & 1.71 &  0.0      & -2.87    & 4.66   & 1.0        & 0.0        &  0.12  \\ 
	P (ads) & Hill & 15.19 & 2.33               & 1.18 &  0.5      & -5.28    & 4.89   &-0.9        &-0.2        & -0.16  \\ 
	\hline                                        
	B (sub) & -    & 15.21 & 1.93               & 0.00 &  0.0      & -6.21    & 5.15   &-1.4        &-0.6        & -0.08  \\ 
	N (sub) & -    & 15.03 & 1.80               & 0.14 &  0.9      & -5.64    & 4.67   &-2.1        &-0.4        & -0.03  \\ 
	Al (sub)& -    & 15.43 & 2.36               & 0.44 &  0.0      & -2.28    & 5.04   & 1.6        & 0.3        & -0.03  \\ 
	P (sub) & -    & 15.24 & 2.24               & 0.47 &  0.0      & -4.84    & 4.55   &-1.2        &-0.1        &  0.02  \\ 
	\hline \hline
\end{tabular}
\end{table}

In Table \ref{table_main}, the electronic properties
of B, N, Al or P adsorbed silicene are also presented.
Both B and Al atoms adsorbed on the valley site
as well as the N atom adsorbed on the bridge site
turn the semimetallic silicene into a nonmagnetic
metal. Adsorption of P atom on the hill site introduces
matallicity with a net magnetic moment of $0.5 \mu_{\mathrm{B}}$.
The work function (measured at the side of the adsorbate)
is slightly lowered to 4.73 eV for B 
and 4.66 eV for Al adsorption from 4.77 eV for intrinsic silicene.
On the other hand, adsorption of a N or P atom increases
the workfunction to 5.02 eV and 4.89 eV, respectively. 

We also present the partial density of states analysis
for B, N, Al and P in Figs.~\ref{fig3}(a-d).
For B adsorption, the metallic bands 
crossing the Fermi level are formed by the hybridization of the B-$p_{xy}$
peaks with the Si-$p_{xy}$ and the Si-$p_{z}$ states.
However, since the adsorption of N occurs on top of a Si-Si bond,
mixing of N states with the silicene states is more complex.
It appears from Fig.~\ref{fig3}(b) that while the main contribution
originates mainly from the hybridization of Si-$p_{z}$ and N-$p_{z}$ bands,
tails of $s$ and $p_{xy}$ states of both silicene
and N play a role in the metallic behavior.
It is also seen that the linearly crossing $p$ bands
of silicene at the K symmetry point are disturbed
because of the interaction with the N-$p_{z}$ states
albeit not completely removed as for B adsorption.
The electronic band dispersion character of Al adsorption
is qualitatively identical to the one of the B adsorption,
caused by the similar adsorption site and valence electron
configuration. The last case of P adsorption
shows the metallic bands predominantly formed by the hybridization
of the P-$p_{xy}$ states with the Si-$p_{xy}$
and the Si-$p_{z}$ states and partially produced
from silicene $s$ states. The linear crossing
of $p$ bands of silicene at the K symmetry point is again
completely removed.

On a graphene lattice, upon B adsorption, 0.4 e
is transferred to graphene and upon N adsorption
0.7 e is transferred from graphene to N.\cite{nakada_2011}
However, due to its $sp^{3}$-like lattice structure,
silicene has a highly reactive surface and therefore
some different adsorption characteristics can be expected.
The charge transfer occurs from silicene
to the adsorbate for the B, N and P adatoms. The opposite
charge transfer is observed for the Al adatom.
The net charge located on B, N, Al and P
adatom has a value of -1.6, -2.0, 1.0 and -0.9 e, respectively.
Moreover, in order to examine the reliability
of these results,
we have also performed a charge population analysis
with the iterative modified Hirshfeld method.\ \cite{hirshfeld_1977,bultinck_2007-ih}
We found that all the methods provide the same
qualitative insight and support for the conclusion
that the character of the charge transfer is driven
by the differences of electronegativity among the
atoms. Silicene donates/accepts electrons
to/from adsorbates with higher/smaller electronegativity.

Substitutional doping of silicene induces an opposite
change of the work function as compared to adsorption.
B or Al substitution increase and N or P substitution
slightly decrease the work function.
The electronic band structure together
with the density of states are shown
in Figs.~\ref{fig3}(e-h).
The Dirac cone composed of $p_{z}$-bands is preserved
for B, Al and P substitution. It has been shifted above the Fermi level
for B and Al substituents and below the Fermi level for
a P atom. It is not altered by the $p_{xy}$ states
of the foreign atom which are present
as deep bands in the final electronic structure.
The character of substitution with a N atom is of a different nature,
the linear crossing of $p$-bands is destroyed and
the metallic bands in the electronic structure are produced
from both $p$ and $s$ states located on N and the surrounding 
Si atoms. These states originate from unoccupied states
of the free N atom and align with the Fermi level in doped silicene.

The calculated charge transfers for the substitutional
doping are of the same magnitude as for adsorption,
with 1.4, 2.1, -1.6 and 1.2 e transferred
from silicene to a B, N, Al and P atom, respectively.
Only N-doped silicene becomes ferromagnetic
with a net magnetic moment of 0.9 $\mu_{\mathrm{B}}$.

The size of the gap as well as the
character of the electronic dispersion bands may be 
altered by the adsorption or substitution
of foreign atoms with the applied spin-orbit interaction
as in the case of graphene decorated by 5d transition-metal
adatoms \cite{zhang_2012-soc_g}.
We performed calculations with spin-orbit coupling
included in order to investigate its effect in all the studied
systems and found no significant differences in the
electronic band dispersion. 
The variances were found to be in the order of
1 meV and virtually vanishing around the Fermi level.
Therefore, we do not report these results in Fig.~\ref{fig3}.
The cause of such negligible coupling is the small diameter
of the adsorbate/substituent atoms and the small or not existing
net magnetic moment of the aforementioned structures.\cite{hu_2012-soc}

\subsection{Stability Analysis}

\begin{figure}
	\includegraphics[width=8.5cm]{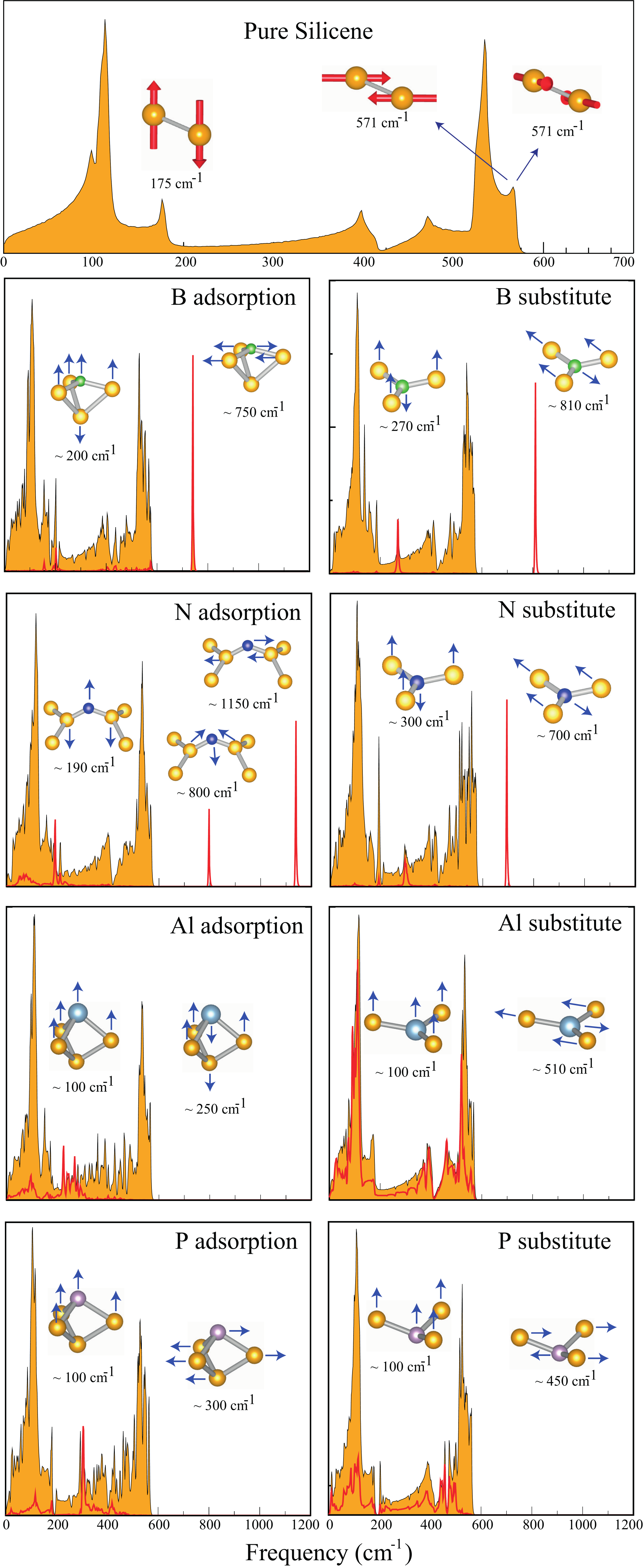}
	\caption{\label{fig-phonons}
	(Color online) Phonon dispersion for adsorption
	and substitution cases of B, N, Al and P on silicene.
	Total DOS of the structures is shown by the filled area 
	(orange color). Projected DOS belonging to
	foreign atoms are represented by lines (red).
	$\Gamma$ point vibrational motions are also depicted.
	}
\end{figure} 

\begin{figure*}
	\includegraphics[width=17cm]{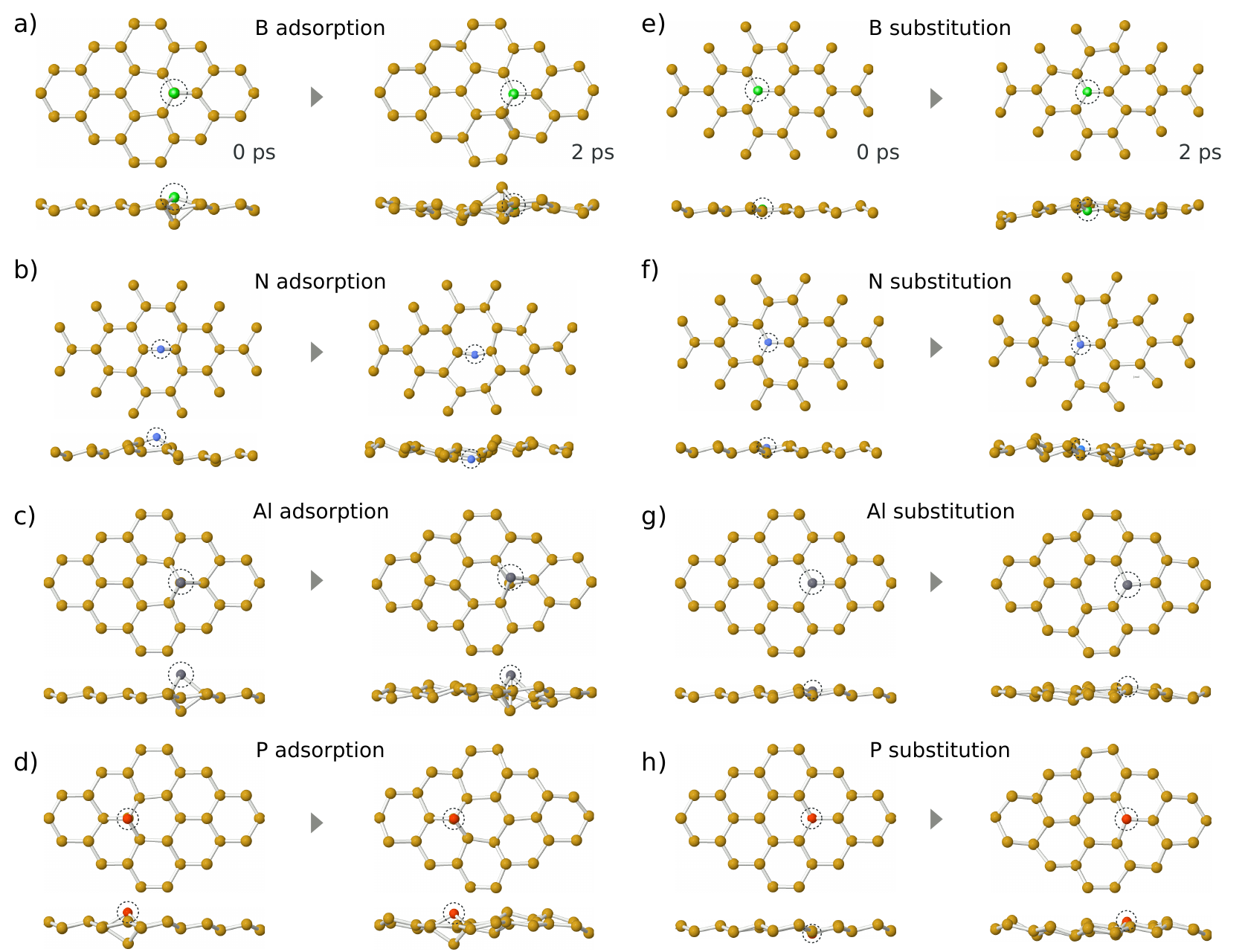}
	\caption{\label{fig-md}
	(Color online) Top and side view of the ground state
	geometry (left) and the changed geometry at $T = 500$ K
	after 2 ps of molecular dynamics (right).
	The adsorbate/substituent atoms are marked
	with an open dotted circle for clarity.
	}
\end{figure*}

Characterization and analysis of the surfaces, crystal structures and molecules can
be done by various spectroscopy techniques. Among these particularly
Raman spectroscopy allows highly accurate detection and analysis of
the vibrational properties of structures even at the nanoscale. Therefore, one can
deduce the stability and characteristic vibrational properties of a foreign atom
adsorbed/substituted silicene layers via their phonon spectrum.
Fig.\ \ref{fig-phonons} shows the evolution of the phonon density of states of
pristine silicene with adsorption and substitution of B, N, Al and P atoms. 
Pristine silicene's phonon DOS comprises three main peaks originating from the
$\Gamma$ atomic motions: an acoustic mode at $\sim$100 cm$^{-1}$,
an out-of-plane optical mode at $\sim$175 cm$^{-1}$ and a degenerate
in-plane optical mode at $\sim$571 cm$^{-1}$. Note
that the peaks at 396, 472 and 535 cm$^{-1}$, are due to the flat slope of the
optical branch at the M-K, M and M symmetry points, respectively. We also present
adsorbate/substituent-projected phonon DOS and related vibrational motions at
the $\Gamma$ point in Fig. \ref{fig-phonons}. 

Adsorption/substitution of first-row elements B and N do not result in a
significant change of the DOS. Due to the coupling of B
and N adsorbates/substituents to the acoustical phonon branch of pristine silicene
several sharp peaks appear between 200 and 300 cm$^{-1}$. Obviously, among the
two adsorbates, B and N, the bridge-site-bonded N adatom is more likely to mix with
silicene's acoustic branch. In addition to these, high frequency
adsorbate/substituent induced modes appear between 700 and 1200
cm$^{-1}$. Our eigenvector analysis reveals that these high energy modes
correspond to in-plane bond-stretching motion of the adatom and the neighboring
silicon atoms.

Consistent with the binding energies calculated in the previous section, the high
frequency in-plane phonon mode of B-substituted silicene takes a higher value than
that of the B-adsorbed one. However, such a comparison is not possible for the N
atom because of the different geometries of the adsorption and substitution
cases. Since the adsorption of a N atom occurs on the
bridge site of two silicon atoms, one can expect vibrational
characteristics different from those of N-doped graphene. While there is no
clear N-induced peak observed in graphene, upon adsorption (substitution) of N
to silicene well separated phonon modes (mode) at $\sim$800 cm$^{-1}$ and
$\sim$1150 (only at 700) cm$^{-1}$ appear. Furthermore, compared with
B-substitution, N substituted silicene's high-frequency mode is softer because
of the larger atomic mass of N. These modes correspond to in-plane and
out-of-plane motion of the N atom and the nearest two Si atoms. Contributions of second
and third nearest neighbor Si atoms are negligible for these high-frequency
modes. Because of their bond-stretching nature, adsorbate- and
substituent-originated optical branches can be expected to be Raman-active
modes. 

The experimentally reported atomic weights (covalent atomic radii) of Si, B, N, Al
and P are 28.09, 10.81, 14.01, 26.98 and 30.97 amu (111, 84, 71, 121 and 107 pm)
and therefore due to the similar properties of second row
elements to Si atoms, for the adsorption/substitution of Al and P atoms
different effects on the vibrational spectra can be expected.
We report a common characteristic for second row elements,
that is the absence of high-frequency bond stretching modes. 
Due to the quite similar atomic weight and radii of Si, Al and
P atoms, their effect on the phonon DOS is almost negligible and may not be
observed by experimental tools. A widely broadened
phonon DOS of Al and P adsorbed silicene, which is different from the narrow
peaks in the B and N adsorption case, indicates the larger coupling of Al atoms to
the silicene lattice. Similarly, the substitution of Al and P does not result in the
presence of high energy modes. It is also seen that while substituent B and
N atoms do not couple with optical modes of silicene, Al and P atoms entirely
contribute to both acoustic and optical modes.

The discussed structures, that are found to be stable in terms
of the total energy optimization and phonon calculations, may not be 
stable at high temperatures. To address questions regarding thermal 
stability of a B and N atom adsorbed/substituted silicene
monolayers we consider the effect of temperature
by employing \textit{ab initio} MD calculations.
MD simulations were performed at a temperature of 500 K
for the structures initially optimized at $T = 0$ K.

In Fig.~\ref{fig-md} geometric structures are presented for
the initial and final state of the MD calculations.
The simulation of B adsorption, as can be seen in Fig.\ \ref{fig-md}(a),
shows that the B atom stays incorporated within the honeycomb structure
but the Si atom, that has been initially pushed down from its position,
has started to move towards the neighboring sites.
During the 2 ps of MD simulations it is attached predominantly
on the hollow and the top sites. However, the adsorbed B atom remains stable.
Similarly, the substituent B atom (Fig.\ \ref{fig-md}(e))
continues to be bonded to the three Si atoms for the whole simulation period.

Although the adsorption geometry for the N atom is different from
B-adsorbed silicene, it remains stable after 2 ps at 500 K.
Even the characteristic out-of-plane motion of the N atom (from
one side to the other one), that can be clearly seen in Fig.\ \ref{fig-md}(b),
is not able to break the strong Si-N-Si bonds.
Similar to the B-substitution case, the N-substituted silicene,
depicted in Fig.\ \ref{fig-md}(f), remains in-plane
bonded to the three neighboring silicon atoms after more than 2 ps at 500 K.

Simulation of Al and P adsorption (Figs.\ \ref{fig-md}(c) and \ref{fig-md}(d))
as well as Al and P substitution (Figs.\ \ref{fig-md}(g)
and \ref{fig-md}(h)) shows high stability of the adsorbate/substituent
atom with no reconfiguration of the Si lattice
in the vicinity of the foreign atoms.

\section{Conclusions}

In summary, we performed an \textit{ab initio} study of atom adsorption
and absorption of B, N, Al and P atoms on silicene and we obtained the geometric, electronic
and vibrational properties. All the final systems show metallic behavior.
The adatoms are strongly bonded with silicene revealing different preferential
adsorption sites for B or N adatoms as compared to graphene.
We find the charge transfer to be consistent with the
effects ascribed to the element electronegativity. B, N and P
atoms behave like acceptors and the Al atom as donor for silicene.
Analysis of the vibrational modes shows that adatoms
and substituents slightly alter the phonon spectrum of
silicene and several adatom/substituent-induced characteristic branches appear,
which may be probed by Raman measurements. Finally, we showed the stability of
B, N, Al and P adsorbed/substituted silicene layers even at high temperatures.
Our work reveals that silicene has a very reactive and functionalizable surface
and it can serve as an important and novel playground for silicene based novel nanoscale
materials.

\begin{acknowledgments}
This work was supported by the Flemish Science Foundation (FWO-Vl) and the Methusalem foundation of the Flemish government. Computational resources were provided by TUBITAK ULAKBIM, High Performance and Grid Computing Center (TR-Grid e-Infrastructure), and HPC infrastructure of the University of Antwerp (CalcUA) a division of the Flemish Supercomputer Center (VSC), which is funded by the Hercules foundation. H. S. is supported by a FWO Pegasus Marie Curie Fellowship.
\end{acknowledgments}

\end{document}